\documentclass[superscriptaddress,secnumarabic,amssymb,amsmath,nobibnotes,aps,prd,showkeys,showpacs,nofootinbib,twocolumn]{revtex4}

\usepackage{graphicx}
\usepackage{bm}
\usepackage{amsmath}
\usepackage{amsfonts}
\usepackage{amssymb}
\usepackage{epstopdf}
\usepackage{natbib}%
\newcommand{\bee}{\begin{equation}}
\newcommand{\eee}{\end{equation}}
\newcommand{\eaa}{\end{eqnarray}}
\newcommand{\baa}{\begin{eqnarray}}
\def\ni{\noindent}

\usepackage{color}

\begin{document}

\title{\Large Notes on microstates, Tsallis statistics \\ and entropic gravity formalism}

\author{Everton M. C. Abreu}\email{evertonabreu@ufrrj.br}
\affiliation{Grupo de F\' isica Te\'orica e Matem\'atica F\' isica, Departamento de F\'{i}sica, Universidade Federal Rural do Rio de Janeiro, 23890-971, Serop\'edica, RJ, Brazil}
\affiliation{Departamento de F\'{i}sica, Universidade Federal de Juiz de Fora, 36036-330, Juiz de Fora, MG, Brazil}
\author{Jorge Ananias Neto}\email{jorge@fisica.ufjf.br}
\affiliation{Departamento de F\'{i}sica, Universidade Federal de Juiz de Fora, 36036-330, Juiz de Fora, MG, Brazil}
\author{Albert C. R. Mendes}\email{albert@fisica.ufjf.br}
\affiliation{Departamento de F\'{i}sica, Universidade Federal de Juiz de Fora, 36036-330, Juiz de Fora, MG, Brazil}
\author{Daniel O. Souza}\email{danieljfos@gmail.com}
\affiliation{Departamento de F\'{i}sica, Universidade Federal de Juiz de Fora, 36036-330, Juiz de Fora, MG, Brazil}


\begin{abstract}
\noindent 
It is an old idea to realize Einstein's equations as a thermodynamical equation of state. Since then, there has been new conjectures to understand gravity from another point of view. 
In this way we can accept that the gravitational field is not an underlying one like an emergent force from other approaches based on the knowledge of relativity, quantum and black holes thermodynamics, and different statistical formalisms.  
One important question concerning this gravity/thermostatistics correspondence is whether the holographic screen could be well defined for a nonrelativistic case of a source mass.  Hence, to understand the actual role of the holographic screen is a very relevant issue.
In this letter we have analyzed the entropy as a function of the holographic screen in some different scenarios.  We have disclosed modified Newtonian dynamics (MOND) from Verlinde's ideas.  Besides, we have calculated some cosmological elements using the same ideas. The results obtained using MOND will guide us to obtain other cosmological results.  
\end{abstract}
\date{\today}

\maketitle
One of the main challenges of modern theoretical physics is to unify the concepts of quantum mechanics and gravitation.   In the literature, it was questioned if this difficulty is caused by the fact that we are really trying to quantize an effective theory.   Another question would be if gravity is not an underlying force.  In its defense we can say that general relativity is an exact theory that describes the dynamics of the objects that comprise our Universe.   Recently, the detection of gravitational waves by LIGO collaboration \cite{ligo} corroborates the predictions of general relativity.  But, in spite of its success, the question about its fundamentability is still on.

There are some theoretical evidences that show the thermodynamical feature of gravity. As examples we have the 
 works of J. D. Bekenstein and S. W. Hawking \cite{bekenstein,hawking} in which the authors connect the laws of black holes to the ones of thermodynamics concerning the creation of particles by a gravitational field.  Or the obtention of Einstein's equations from entropy proposed by Jacobson  \cite{jacobson}, where he proposed that gravitation must be an effective theory.   Besides, we can mention the works of van Raamsdonk {\it et al} \cite{lmr} where the Einstein equations can be derived from the entanglement's laws. Jacobson established the connection between entropy in \cite{jacobson}, where he derived the Einstein's equations.   He used Clausius relation $\delta E \,=\, T\,\delta S$ and the concept that the matter present can be considered as a part of energy.

At the quantum level, we can understand the spacetime geometry as the entanglement structure of the microscopic quantum state, from where gravity emerges depicting the change in entanglement resulting from matter.   Namely, gravity emerges from the viewpoint of quantum information.   The better way to understand these new ideas is to consider an anti-de Sitter space, where the description of a dual CFT permits one to obtain the microscopic entanglement in a well constructed setting \cite{verlinde2}.

Recently, the work of E. Verlinde \cite{Verlinde} has derived Newton's law of gravity by using holographic arguments.   He has used Bekenstein-Hawking entropy-area relation for black holes.   Verlinde suggested that gravity is an entropy manifestation.   The gravitational force results from changes of information entropy which would be stored in an holographic sphere.  Verlinde considered entropy as the information relative to the position of material bodies around a point mass $M$ at a distance $R$.   Besides, all points at this distance are useful to define a sphere $S$ embedded in a $D=3$ space.   To sum up, when we change the bits of information that are localized on the surface, a force appears as a reaction to that change.   It will be shown in equation \eqref{force3} below.

The formalism of modified Newtonian dynamics (MOND) was constructed by M. Milgrom \cite{mond}, to explain the observed general properties of galaxies such as the velocities of stars in galaxies.  The observed values are larger than the expected ones calculated using Newtonian mechanics.  In MOND, for extremely small acceleration, we would have a violation of Newton's laws.   These small accelerations are a characteristic property of these objects that dwell far from our solar system.   However, a cosmological model based on MOND's concepts has not yet been constructed.   In this work we have used Verlinde's ideas embedded in Tsallis nonextensive statistics to derive MOND's concepts among other results.

\bigskip

Tsallis' statistics \cite{tsallis}, which is an extension of Boltzman-Gibbs's (BG) statistical theory, defines a nonadditive entropy as
\begin{eqnarray}
\label{nes}
S_q =  k_B \, \frac{1 - \sum_{i=1}^W p_i^q}{q-1}\;\;\;\;\;\;\qquad \Big(\,\sum_{i=1}^W p_i = 1\,\Big)\,\,,
\end{eqnarray}

\ni where $p_i$ is the probability of a system to exist within a microstate, $W$ is the total number of configurations and 
$q$, known in the current literature as being the Tsallis parameter or NE  parameter, is a real parameter which measures the degree of nonextensivity and takes different values for each system. 
The definition of entropy in Tsallis statistics carries the standard properties of positivity, equiprobability, concavity and irreversibility. This approach has been successfully used in many different physical system. For instance, we can mention the Levy-type anomalous diffusion \cite{levy}, turbulence in a pure-electron plasma \cite{turb} and gravitational systems \cite{sys}.
It is noteworthy to affirm that Tsallis thermostatistics formalism has the BG statistics as a particular case in the limit $ q \rightarrow 1$ where the standard additivity of entropy can be recovered. 

In the microcanonical ensemble, where all the states have the same probability, Tsallis entropy reduces to \cite{te}
\begin{eqnarray}
\label{micro}
S_q=k_B\, \frac{W^{1-q}-1}{1-q},
\end{eqnarray}
where, at the limit $q \rightarrow 1$, we can recover the usual Boltzmann entropy formula, $S=k_B\, \ln {W}$\,\,.

To begin our work, let us consider a number of microstates $W$, scaling like a volume in the scenario of Tsallis' entropy, so that $W$ can be written as in \cite{MR} such that
\begin{eqnarray}
\label{volume}
W=b  \left(\frac{A}{l_p^2}\right)^\frac{3}{2},
\end{eqnarray}

\ni where $b$ is a dimensionless constant, $A$ is the area of the holographic screen and $l_p$ is the Planck's length in the nonextensive entropy, Eq. (\ref{micro}).  The term in Eq. \eqref{volume} was calculated through LQG considerations at \cite{lt} as a correction for the entropy.  Eq. \eqref{volume} is the volume conrrection to the area law, which is also motivated by a model for the microscopic degrees encompassing the black hole entropy in LQG.   Concerning our work specifically, the important feature is that this term originates the $1/R$ correction term to Newton's laws in Modified Newton's Dynamics (MOND) as an explanation for the registered anomalous galactic rotation curves.   This idea is the bridge that allows us to use Tsallis nonextensive statistics.  It can be clearer soon.  
 
The objective here is to use Eq. \eqref{volume} in the definition of entropy in Eq. \eqref{micro} and to analyze the effect of $q$-parameter and its cosmological effects.   Notice that for $q=0$ we recover the volume correction term to Newton's laws, which confirms the coherence to take the volume term to generalize the number opf microstates $W$, in the particular case of Tsallis entropy.



To begin with, let us review that Verlinde \cite{Verlinde} has stated that the entropy $\Delta S$ of an holographic screen connected to a test particle of mass $m$ moving by a distance $\Delta x$ orthogonal to the screen can be written as $$\Delta S\,=\,2\pi k_B \frac{mc}{\hbar} \Delta x\,\,,$$ which shows that the entropy gained is proportional to the information loss of the test particle, where $\lambda_m = \hbar/mc$ is the Compton wavelength and we can write $\Delta S = 2\pi k_B \Delta x/\lambda_m$.   The general expression of the force which is governed by the usual thermodynamic equation is
\begin{eqnarray}
\label{egen}
F=T \frac{\Delta S}{\Delta x}=T \frac{dS}{dA} \frac{\Delta A}{\Delta x}\,\,,
\end{eqnarray}

\ni where $A=4\pi R^2$ is the area of the holographic screen. Suppose we have two masses, one is a test mass $m$ and the other, $M$, considered as a source. The holographic screen will be centered around the source mass $M$. The energy of the holographic screen is given by
\begin{eqnarray}
\label{energy}
E = M c^2\,\,.
\end{eqnarray}

\ni We will use  that the bits of information scale proportionally to the area of holographic screen as \cite{MR}
\begin{eqnarray}
\label{area}
A=Q N\,\,,
\end{eqnarray}

\ni where $N$ is the number of bits and the constant $Q$, the fundamental charge \cite{MR}, will be determined later. The total energy of the bits on the screen is given by the equipartition law of energy
\begin{eqnarray}
\label{equipar}
E = \frac{1}{2} N k_B T\,\,.
\end{eqnarray}

When the test mass $m$ is at a distance 
\begin{eqnarray}
\label{distance}
\Delta x = \eta \lambda_m\,\,,
\end{eqnarray}

\ni away from the holographic surface $S$, the entropy
of the surface modifies by one fundamental unit $\Delta S$ fixed
by the discrete spectrum of the area of the surface,
where $\lambda_m=\hbar/m c$ is the Compton wavelength. And the entropy gradient points radially
from the outside of the surface to the inside, as can be seen from 
\bee
\label{}
\Delta S \,=\, \frac{\partial S}{\partial A}\, \Delta A\,\,.
\eee  

Then, from Eq. (\ref{area}) we have that
\begin{eqnarray}
\label{darea}
\Delta A = Q\,\,,
\end{eqnarray}

\ni where it was assumed that $\Delta N=1$. Combining Eqs. (\ref{energy})-(\ref{distance}) and (\ref{darea}),   we have that
\begin{eqnarray}
\label{force2}
F = \frac{G M m}{k_B r^2} \, \frac{Q^2}{2\pi \eta l_p^2}\, \frac{dS}{dA}\,\,.
\end{eqnarray}

\ni Defining conveniently that $Q^2=8\,\pi\, \eta\, l_p^4$, we can write that
\begin{eqnarray}
\label{force3}
F = \frac{G M m}{r^2} \, 4 \frac{l_p^2}{k_B} \, \frac{dS}{dA}\,\,.
\end{eqnarray}

\ni Notice that this expression is enoughly general in order to permit the consideration of any kind of entropies, which is our purpose at this point.
If we use the well known particular relation from black hole entropy $S=k_B A/4 l_p^2$, the Bekenstein-Hawking (BH) formula, where in this case $A$ is the area of the event horizon, i.e., a screen that sets the point of no return. Quanticaly speaking, a black hole creates and emits particles as if it was a black body with temperature $T$ \cite{hm}.

It can be shown that we can obtain the usual Newton law of gravitation, $F=G M m/r^2$ from Eq. (\ref{force3}).  Newton's gravitational theory just states how the law works but, however, it does  not tell us why they work.  

Using (\ref{micro}) and \eqref{volume}  into (\ref{force3}) we obtain a modified Newton's law of gravitation written as
\begin{eqnarray}
\label{force4}
F = \frac{G M m}{r^2} \, 6 b^{1-q} \, \left(\frac{A}{l_p^2}\right)^\frac{1-3q}{2}\,\,.
\end{eqnarray}

\ni We can observe that when we make $q=\frac{1}{3}$ in above equation we recover the usual Newton law of gravitation if $b=(\frac{1}{6})^\frac{3}{2}\approx 0.07$.   Substituting relation (\ref{volume}) into (\ref{micro}) it is direct to see that
\begin{eqnarray}
\label{entropyn}
S=k_B \frac{b^{(1-q)} \left(\frac{A}{l_p^2}\right)^{\frac{3}{2}(1-q)}-1}{1-q}\,\,,
\end{eqnarray}

\ni which gives us a kind of ``master expression" for the entropy as a function of the nonextensive parameter and the $b$-parameter. This last one can be adjusted conveniently as will be seen just below.  This parameter can provide us the results for the black hole and for Verlinde's holographic screen.
For example, using $q=\frac{1}{3}$ and $b=(\frac{1}{6})^\frac{3}{2}$ in (\ref{entropyn}) we obtain
\begin{eqnarray}
\label{botta} 
S=k_B \frac{A}{4l_p^2}\,\,, 
\end{eqnarray}

\ni which is the BH formula.  Namely, in Newton's gravitational scenario we can have the black hole entropy.  One can think that this result connects the thermodynamical Bekenstein-Hawking formula for black holes with the classical Newton's expression for gravity, which could suggest a thermodynamical emergent gravitation.
On the other hand, Botta Cantcheff  and Nogales \cite{BN} have shown that we can derive the usual entropy of black holes by using a volume microstates scaling law (\ref{volume}) and the Tsallis' nonextensive 
entropy (\ref{micro}). 

For $q=1$ (BG) scenario and same $b$ we have that
\bee
\label{15}
F_{q=1}\,=\,\frac{GMm}{r^2} \frac{6 \,l^2_p}{k_B A}\,\,,
\eee

\ni which of course is not Newton's second law.   

\ni For $q=1/3$ in \eqref{force4}, we recover the usual Newton law of gravitation if $b=0.68$, as we said before.




From \eqref{entropyn}, as we have said before, for $q \rightarrow 1$ we have the BG statistics but at the same time it is a divergence point, so, after calculating the limit 
\baa
\label{17}
&&S_{q\rightarrow 1}\,=\,\lim_{q\rightarrow 1} k_B\frac{b^{1-q}\Big(\frac{A}{l^2_p}\Big)^{\frac 32 (1-q)}-1}{1-q} \nonumber \\
\mbox{} \nonumber \\
&&\Longrightarrow \quad S_{q \rightarrow 1} \,=\,k_B\; ln \,\bigglb[b\bigglb(\frac{A}{l^2_p}\biggrb)^{3/2}\biggrb]  \label{18} \\
&&\Longrightarrow \quad W\,=\,b\bigglb(\frac{A}{l^2_p}\biggrb)^{3/2} \label{19}\,\,, 
\eaa

\ni which confirms the expression in \eqref{volume}.


\ni which keeps the nondimensional property of $b$.

Now, from \eqref{entropyn}, i.e., using the entropy as the starting point, we can compute that
\bee
\label{21}
\frac{dS}{dA}\,=\,\frac 32 k_B \frac{b^{1-q}\,A^{\frac 12 (1-3q)}}{(l_p^2)^{\frac 32 (1-q)}}\,\,.
\eee

Substituting this result into Eq. \eqref{force3} we have that
\bee
\label{22}
F\,=\,\frac{6GMm}{r^2} \frac{b^{1-q}\,A^{\frac 12 (1-3q)}}{(l_p^2)^{\frac 12 (1-3q)}}\,\,, 
\eee

\ni where, in order to have the second law, we can write that
\bee
\label{23}
b\,=\,\Bigglb[ \frac 16 \frac{(l_p^2)^{\frac 12 (1-3q)}}{A^{\frac 12 (1-3q)}} \Biggrb]^{\frac{1}{1-q}}\,\,,
\eee

\ni which gives us the result $q=1/3$ since $b$ is a number, confirming the Newton's law.   


For the BG regime, $q=1$, into Eq. \eqref{22} we have 

\bee
\label{AA}
F_{q=1}\,=\,\,\frac{GMm}{r^2}\,\frac{6}{A/l^2_p}\,\,.
\eee

\ni where $A=4\pi r^2$.  Using $q=0$ in (\ref{force4})  we have that

\begin{eqnarray}
\label{forcem}
F &=& \frac{G M m}{R^2} \, 6 b \, \left(\frac{A}{l_p^2}\right)^\frac{1}{2}\nonumber\\
&=& \frac{12 \pi^\frac{1}{2} b}{l_p}\,\frac{G M m}{R}\,\,,
\end{eqnarray}

\ni which is just the Newtonian force established by Modified Newtonian Dynamics (MOND) approach \cite{mond}. From the rotational movement of the galaxies we have that 
$v^2=\sqrt{G M a_0}$ and substituting this velocity into \eqref{forcem} we have that

\begin{eqnarray}
\label{29}
\frac{v^2}{R}=\frac{12 \sqrt{\pi} b}{l_p}\,\frac{G M}{R}\,\,,
\end{eqnarray}

\ni hence

\begin{eqnarray}
\label{31}
b\,=\,\frac{v^2 l_p}{12\sqrt{\pi}GM}\,=\,\frac{l_p}{12}\sqrt{\frac{a_0}{\pi GM}}\,\,, 
\end{eqnarray}

\ni which is a viable result for $b$.   Let us analyze other consequences of this $b$-value, which reproduces MOND, as we saw above.

From Eq. \eqref{force2} we have that,
\bee
F = \frac{G M m}{k_B r^2} \, \frac{Q^2}{2\pi \eta l_p^2}\, \frac{dS}{dA}\,\,, \nonumber
\eee

\ni which is the thermodynamical expression for the Newtonian force.   So,

\begin{eqnarray}
\label{f1}
m \ddot{r}_H\,=\,m \ddot{a} r_0 = \frac{G M m}{a^2 r^2} \, \frac{4 l_p^2}{k_B}\, \frac{dS}{dA}
\end{eqnarray}

\ni where $r_H$ is the apparent horizon, i.e., $r_H = a r_0$, and 

\begin{eqnarray}
\label{f2}
\Longrightarrow  \ddot{a}=\frac{GM}{k_B a^2r^3} \, \frac{4 l_p^2}{k_B} \,\frac{dS}{dA}\,\,,
\end{eqnarray}

\ni which, based on \cite{padmanabhan5}, the acceleration in Eq. \eqref{f2} results from the active gravitational mass, which is the well known Tolman-Komar mass \cite{tk} given by

\begin{eqnarray}
\label{mf}
M=(\rho +3p) \frac{4\pi}{3} a^3 r^3\,\,,
\end{eqnarray}

\ni which is proportional to the scale function.  Substituting Eq. \eqref{mf} into Eq. \eqref{f2} we have that
\begin{eqnarray}
\label{f3}
\frac{\ddot{a}}{a}=-\frac{16\pi}{3} G (\rho + 3p) \frac{l_p^2}{k_B} \,\frac{dS}{dA}\,\,,
\end{eqnarray}

\ni and from this last equation, by multiplying sides of this last equation by $\dot{a} a$, we have that
\bee
\dot{a}\,\ddot{a}\,=\,-\,\frac{16\pi}{3}\frac{d}{dt}(\rho\,a)\,\frac{l_p^2}{k_B}\frac{dS}{dA}\,\,,
\eee

\ni and integrating both sides we have that
\begin{eqnarray}
\label{f4}
H^2+\frac{k}{a^2}=\frac{32\pi G}{3k_B} \, \frac{l_p^2}{a^2} \, \int d(\rho a^2) \,\frac{dS}{dA}\,\,,
\end{eqnarray}

\ni which is an entropic version of the Friedmann equation and where we have used the continuity equation
\begin{eqnarray}
\label{cont}
\dot{\rho}+3H (\rho+p)=0\,\,.
\end{eqnarray}

So, to calculate the active mass, the TK mass, we have to solve the differential equation in \eqref{f2}, which can be written as
\bee
\ddot{a}\,=\,\frac{GM}{a^2 r^3_0} \frac{4 l_p^2}{k_B}\,\frac{dS}{dA}\,=\,6\,\frac{GM}{k_B a^2 r^3_0} \,b^{1-q} \Big(\frac{A}{l^2_p} \Big)^{\frac 12 (1-3q)}\,\,,
\eee



\ni where we have used Eq. \eqref{entropyn}. Hence, for $A=4\pi r_H^2=4\pi a^2 r_0^2$, we have that
\bee
\label{37.1}
\ddot{a}\,=\,\frac{6GM}{k_B a^{1+3q}r^{2+3q}_0} b^{1-q} \Big( \frac{4\pi}{l^2_p}\Big)^{\frac 12 (1-3q)}\,\,,
\eee

\ni where $r_H$ is the apparent horizon, $r_H\,=\,1/H$.   For $q=1$ (the BG limit) we have that $$\ddot{a}\,=\,\frac{3GMl^2_p}{2\pi k_B a^4 r^5_0}\,\,.$$ which is also a non-linear differential equation with numerical solution for the scale factor.   Numerical computation is out of the scope of this letter.


Another fundamental relation in cosmology is the continuity equation
\begin{eqnarray}
\label{cont2}
\dot{\rho}+3H (\rho+p)=0\qquad\,\,, 
\end{eqnarray}

\ni and using the state equation $p=\omega \rho$ we have that
\baa
\label{43}
&&\dot{\rho}\,+\, 3\,H \rho\,(1+3\omega)\,=\,0 \nonumber \\
\mbox{} \nonumber \\
&\Longrightarrow& \rho\,=\,\frac{\rho_0}{a_0} a\,e^{-3(1+\omega)}\,\,,
\eaa

\ni where $\omega$ is assumed to be constant, which means a non-adiabatic model of DE, for example.   This last equation 
means that we can obtain a differential equation for the scale factor as a function of the equation of state and the $q$-parameter.   Having said that, we can obtain a master equation for the scale factor.   Let us use Eq. \eqref{37.1} with the solution \eqref{43}

\bee
\ddot{a}\,=\,8\pi \frac{G}{a^{3q}}\frac{\rho}{\rho_0}\,e^{-3(1+\omega )}\,b^{1-q}\bigglb( \frac{4\pi r^2_0}{l^2_p} \biggrb)^{\frac 12 (1-3q)}\,\,.
\eee

\ni Hence, for $q=0$ we have MOND considerations and the scale factor is given by

\bee
\ddot{a}\,=\,16\,\pi^{3/2} \frac{G\,r_0 \,\rho\, b}{l_p \rho_0}\, e^{-3(1+\omega)}\,\,,
\eee

\ni for $q=1$ we have the BG case, and the scale factor is

\bee
\ddot{a}\,=\,2\,\frac{G\,\rho\,l_p^2}{a^3\,\rho_0\,r_0^2}\,e^{-3(1+\omega)}\,\,.
\eee

We have seen before that, for the BH entropy relation we have that $q=1/3$ and $b=(1/6)^{3/2}$, so

\bee
\ddot{a}\,=\,\frac{4\pi}{3} \frac{G\,\rho}{a\,\rho_0}\,e^{-3(1+\omega)}\,\,,
\eee

\ni which is the equation for the scale factor for a BH scenario.

In this letter we have explored the role of the holographic screen under the point of view of Tsallis thermostatistics.   We have saw through the volume correction term of the entropy \cite{MR}, which is connected to MOND ideas, that the $q$-parameter can be fixed since viable cosmological issues were considered.   We have found that the $b$-parameter used in \cite{MR} can be also fixed according the physical scenario.

\section{Acknowledgments}

\ni E.M.C.A.  thanks CNPq (Conselho Nacional de Desenvolvimento Cient\' ifico e Tecnol\'ogico), Brazilian scientific support federal agency, for partial financial support, Grants numbers 302155/2015-5 and 442369/2014-0 and the hospitality of Theoretical Physics Department at Federal University of Rio de Janeiro (UFRJ), where part of this work was carried out.  



\begin{thebibliography}{99}

\bibitem{ligo}   B. P. Abbott (LIGO Scientific Collaboration and Virgo Collaboration) et al.,  
  Phys. Rev. Lett. 116 (2016) 061102.

\bibitem{bekenstein}   J.D. Bekenstein,  Phys. Rev. D 7, 2333 (1973);  Phys. Rev. D 9 (1974) 3292.

\bibitem{hawking}   S.W. Hawking, Commun. Math. Phys. 43 (1975) 199;
Phys. Rev. D 14, 2460 (1976); ``The Information Paradox for Black Holes," arXiv: 1509.01147.

\bibitem{jacobson}   T. Jacobson, Phys. Rev. Lett. 75 (1995) 1260;  Phys. Rev. Lett. 116 (2016) 201101; \\
C. Eling, R. Guedens, and T. Jacobson, Phys. Rev. Lett. 96   (2006) 121301; \\
R. Guedens, T. Jacobson, and S. Sarkar, Phys. Rev. D 85 (2012)   064017.

\bibitem{lmr}   N. Lashkari, M.B. McDermott, and M. Van Raamsdonk, JHEP 1404 (2014) 195; \\
T. Faulkner, M. Guica, T. Hartman, R.C. Myers, and M. Van Raamsdonk,  JHEP 1403 (2014) 051;\\
B. Swingle, and M. Van Raamsdonk, ``Universality of Gravity from Entanglement," arXiv: 1405.2933 [hep-th].


\bibitem{verlinde2}   E. Verlinde, SciPost Phys. 2 (2017) 016.

\bibitem{Verlinde} E. Verlinde, JHEP 1104 (2011) 029.

\bibitem{mond} M. Milgrom, Astrophys. J. 270 (1983) 371; {\it ibid} 270 (1983) 365; {\it ibid} 270 (1983) 384.



\bibitem{tsallis} C. Tsallis, J. Stat. Phys. 52 (1988) 479;\\ C. Tsallis, ``Introduction to Nonextensive Statistical Mechanics: Approaching a Complex World," Springer (2009);\\ C. Tsallis, Braz. J. Phys. vol. 29 (1999) 1.


\bibitem{levy} P. A. Alemany and D. H. Zanette, Phys. Rev. Lett. 75 (1995) 366.

\bibitem{turb} C. Anteneodo and C. Tsallis, J. Mol. Liq. 71 (1997) 255.

\bibitem{sys} C. Tsallis, Chaos, Soliton and Fractals 13 (2002) 371;\\ R. Silva and J. S. Alcaniz, Physica A 341 (2004) 208. 

\bibitem{te} C. Tsallis, Chaos, Soliton and Fractals 6 (1995) 539.


\bibitem{MR} L. Modesto and A. Randono, ``Entropic corrections to Newton's law," arXiv: 1003.1998.

\bibitem{lt}   E. R. Livine and D. R. Terno, Nucl. Phys. B 794 (2008) 138.


\bibitem{BN} M. Botta Cantcheff and J. A. C. Nogales, Int. J. Mod. Phys A 21 (2006) 3127.











\bibitem{ananias} J. Ananias Neto, Physica A 391 (2012) 4320.






\bibitem{nos}    E. M. C. Abreu, J. Ananias Neto, E. M. Barboza and R. C. Nunes, EPL, 114 (2016) 55001.

\bibitem{botta}   M. Botta Cantcheff and J. A. C. Nogales, Int. J. Mod. Phys. A21 (2006) 3127. 

\bibitem{hm}  X-G. He and B-Q. Ma,  Chin. Phys. Lett. 27 (2010) 070402.

\bibitem{padmanabhan5}   T.  Padmanabhan, Class. Q. Grav. 21 (2004) 4485.

\bibitem{tk}  R. C. Tolman and Phys. Rev. 35 (1930) 875;\\ A. Komar, Phys. Rev. 113 (1959) 934.


\end{thebibliography}
\end{document}